\documentclass[pra,aps,showpacs,twocolumn]{revtex4}
\usepackage{amsmath}
\usepackage{graphicx}
\usepackage{dcolumn}
\usepackage{bm}

\begin{document}
\title{Cavity QED with cold atoms trapped in a double-well potential}
\author{Jiang-Ming Zhang, Wu-Ming Liu, and Duan-Lu Zhou}
\affiliation{Beijing National Laboratory for Condensed Matter
Physics, Institute of Physics, Chinese Academy of Sciences, Beijing
100080, China.}

\begin{abstract}
We investigate the interplay dynamics of a Cavity QED system, where
the two-level atoms are trapped in a double-well potential, and the
cavity mode, with a frequency largely detuned to the atomic level
splitting, is driven by a probe laser. The interaction between the
center-of-mass motion of the atoms and the cavity mode is induced by
the position dependent atom-field coupling. The dynamics of the
system is characterized by two distinct time scales, the inverse of
the atomic interwell tunneling rate and the inverse of the cavity
loss rate. The system shows drastically different (quasi) steady
behaviors in the short-time and long-time intervals.
\end{abstract}
 \maketitle

\section{Introduction}
\label{intro} The past decade has witnessed great advances in both
the fields of cold atoms and cavity quantum electrodynamics
(Cavity-QED), and the overlap between the two fields is
ever-growing. A remarkable achievement in this direction is the
successful coupling of a Bose-Einstein condensate to a quantized
field mode of a high-finesse optical cavity \cite{esslinger nature,
tonature}. Besides that, deterministic loading of individual atoms
in a micro-cavity is demonstrated \cite{chapman} and submicron
positioning of single atoms in the cavity is achieved \cite{chapman,
nufmann}, which allows control of the atom-field coupling via its
position-dependence.

Theoretically, Mekhov, Maschler, and Ritsch proposed to probe the
superfluid-insulator transition of cold atoms in optical lattices by
the transmission spectra of an optical cavity \cite{Ritsch_nature}.
The atoms couple to a quantized cavity mode dispersively and hence
act as some moving refractive media in the cavity. The cavity
transmission spectra directly reflects the quantum or classical
distributions of the atoms, which characterize the superfluid or
insulator phases respectively. This non-destructive proposal
exploits the fact that in the domain of strong coupling, even one
atom is enough to shift the cavity resonance significantly.
Techniques based on this knowledge have been developed to detect the
existence of atoms in a cavity \cite{ye}, and most recently, been
employed to study the correlation, statistics and dynamics of
matter-wave fields \cite{esslinger}.

From the point of view of atomic optics and quantum information,
Ref.~\cite{Ritsch_nature} also provides us with a model of rich
coupled atom-field dynamics \cite{optics comm}. The atoms
effectively influence the field dynamics by shifting the resonance
of the field mode, while in turn the field intensity determines the
dipole potential for the atoms. The former effect is essential for
the result of Ref.~\cite{Ritsch_nature} and is treated in detail.
However, the latter effect is neglected. The atomic dynamics is
avoided by prescribing a state (phase) for the atoms. Furthermore,
the interaction and the coupling to the environment may induce
entanglement between the atomic and field subsystems, and cause
decoherence of the subsystems, respectively. All these aspects of
the system are rarely investigated in Ref.~\cite{Ritsch_nature}.

The purpose of this paper is to investigate the dynamics of the
composite atom-field system, with the emphasis on the interplay
between the two sides, the correlation and the entanglement between
them. We shall consider a ``two-site version'' of the model
presented in Ref.~\cite{Ritsch_nature}. Atoms are trapped in a
double-well potential and interact dispersively with a damped and
driven field mode. The two traps are placed asymmetric to the field
mode so that the atomic tunneling dynamics is coupled to the field
dynamics. Under the two-mode approximation, the freedoms of the
atoms are reduced to minimum and can be taken into full account. To
gain insight into the dynamics of the system, we assume the system
starts from an initial state and evolves towards the steady state.
We find that this process involves two distinct time scales, one is
the atomic tunneling rate and the other the cavity loss rate, with
the latter much faster than the former. These two incommensurate
time scales lead to distinct temporal structures of the dynamics. In
the short-time interval, where the atomic tunneling can be
neglected, it is found that the model we consider is analogous to
the Dicke model in the dispersive regime, for which a good
understanding exists \cite{klimov}. Detailed analytical results are
obtained and, by the way, the main result of
Ref.~\cite{Ritsch_nature} is recovered. In the long-time interval,
however, the atomic tunneling plays an important role. Strong
population transfer between different atomic states is observed, and
the system displays substantially different behavior than in the
short-time interval.

This paper is organized as follows. In Sec.~\ref{sec 2} the basic
model is introduced and the Hamiltonian of the atom-field system is
derived under the two-mode approximation. Then in Sec.~\ref{sec 3},
based on the master equation, the short-time and long-time behaviors
are investigated both analytically and numerically. Finally, our
results are summarized in Sec.~\ref{sec 4}. The connection with the
Dicke model is discussed in Appendix A and some useful inequalities
are derived in Appendix B.

\section{Basic model and the hamiltonian}
\label{sec 2} In this work, we consider the combination of a
double-well and an optical cavity, two paradigm models in physics.
We assume $N$ two-level bosonic atoms with mass $m$ and transition
frequency $\omega_a$ are trapped in a double-well potential $V(x)$
and loaded in an optical cavity, where they interact with a cavity
field mode with frequency $\omega_c$. The cavity is coherently
pumped through the mirror by a weak laser with frequency $\omega_p$
and amplitude $\eta$. We also assume the atom-field detuning is much
larger than the atomic spontaneous emission rate and the Rabi
frequency. Under this condition, the atomic upper level can be
adiabatically eliminated \cite{meystre,ritsch prl}, i.e., the atomic
internal dynamics is neglected.

After adiabatic elimination of the atomic upper state, the
single-atom-plus-field Hamiltonian in the frame rotating at the
frequency of the pumping field is~\cite{ritsch prl}:
\begin{equation}\label{H0=Ha+Hphoton}
    H_0=H_{ph}+H_s,
\end{equation}
where $H_{ph}$ is the rotating frame Hamiltonian for the driven
field,
\begin{equation}\label{Hphoton}
    H_{ph}=-\Delta a^\dagger
a+\eta(a+a^\dagger),
\end{equation}
with $\Delta=\omega_p-\omega_c$ being the pump-cavity detuning, and
\begin{equation}\label{Ha}
    H_s=\frac{p^2}{2m}+V(x)+u^2(x)(U_0a^\dagger a),
\end{equation}
which is the Hamiltonian for a single atom in the superposition of
the classical potential $V(x)$ and the quantum potential
 $u^2(x)U_0a^\dagger a$~\cite{ritsch prl}. Here $u(x)$ is the field mode function
with its magnitude at the antinode normalized to unity. The
parameter $U_0=g_0^2/(\omega_c-\omega_a)$, with $g_0$ being the
atom-field coupling at the antinode.

The many-atom-plus-field Hamiltonian, taking into account the direct
interaction between the atoms which is characterized by the $s$-wave
scattering length $a_s$, is:
\begin{eqnarray}\label{many particle hamiltonian}
    H&=&H_{ph}+\int d^3x \Psi^\dagger(x)H_a\Psi(x)  \nonumber \\
    & & +\frac{1}{2}\frac{4\pi a_s}{m}\int d^3x
    \Psi^\dagger(x)\Psi^\dagger(x)\Psi(x)\Psi(x),
\end{eqnarray}
where $\Psi(x)$ is the atomic field operator and we take
$\hbar\equiv1$ here and henceforth. Under the two-mode approximation
for the atomic freedoms, the validity of which has been well
established Ref.~\cite{d f walls, jaksch}, the atomic field operator
$\Psi(x)$ has two contributions:
\begin{eqnarray}
\label{a field operator}
\Psi(x)&=&b_1 w_1(x)+b_2 w_2(x) \nonumber   \\
  &=& b_1 w(x-x_1)+ b_2 w(x-x_2),
\end{eqnarray}
Here we assume a symmetric double well with the two minima at $x_1$
and $x_2$ . The two modes $w_1(x)$ and $w_2(x)$ are localized in the
left and right traps respectively, and satisfy the orthonormal
relation $\int d^3x w_i^\ast(x) w_j(x) = \delta_{ij}$, $(i,j)=1,2$.
The operator $b_i^\dagger$ ($b_i$) ($i=1,2$) creates (annihilates)
an atom in the mode $w_i(x)$. Substituting Eq.~(\ref{a field
operator}) into Eq.~(\ref{many particle hamiltonian}), and keeping
only terms with dominating contributions, we obtain
\begin{equation}\label{h=h1+h2+h3}
    H=H_{ph}+H_a+H_{int},
\end{equation}
where $H_a$ is the Hamiltonian for the atomic subsystem,
\begin{equation}\label{h1}
    H_a= -t(b_1^\dagger b_2+b_2^\dagger
    b_1)
    +\frac{u}{2}(n_1(n_1-1)+n_2(n_2-1)).
\end{equation}
Here we introduce the atom number operators $n_i=b_i^\dagger b_i$ $
(i=1,2)$ and drop the term associated with the zero-point energy.
The atomic tunneling rate $t$ and the on-site interaction energy $u$
are defined as
\begin{eqnarray}
-t &=& \int d^3x
    w_1^\ast(x)(-\frac{\bigtriangledown^2}{2m}+V(x))w_2(x),  \\
u&=&\frac{4\pi a_s}{m}\int d^3x |w_{1,2}(x)|^4.
\end{eqnarray}
$H_{int}$ describes the interaction between the atoms and the field,
\begin{eqnarray}\label{h2}
    H_{int}&=&\int d^3x
    \Psi^\dagger(x)u^2(x)\Psi(x)(U_0 a^\dagger a)
    \nonumber \\
    &\simeq&(J_1n_1+J_2n_2)(U_0a^\dagger a),
\end{eqnarray}
where the dimensionless coefficients $J_{1,2}$ are defined as
\begin{equation}
J_{1,2} = \int d^3x u^2(x) |w_{1,2}(x)|^2,
\end{equation}
which reflect the overlap between the atomic modes and the field
mode. Note that $J_{1,2}$ are bounded, $0\leq J_{1,2} \leq1$, which
follows from the normalization conditions of $u(x)$ and
$w_{1,2}(x)$. If the field mode $u(x)$ varies slowly in the range of
the spread of the atomic modes, we can take the ``tight confinement
approximation''~\cite{stanper kurn, Ritsch_nature} $J_i\simeq
u^2(x_i)$ ($i=1,2$). It is clear from Eq.~(\ref{h2}) that the
interaction between the atoms and the field is twofold. For the
atoms, the depths of the two traps are shifted while for the field
the energy per photon is renormalized.

We shall discriminate two different cases: (\romannumeral 1)
$J_1=J_2$; (\romannumeral 2) $J_1\neq J_2$. The former case is
trivial, because in this case the dynamics of the atoms and the
field is decoupled, the field is indifferent to the distribution of
the atoms in the two traps. Thus we concentrate on the case $J_1\neq
J_2$. Without loss of generality, we assume $J_1=1$, $J_2=0$. This
is always reasonable mathematically because we can define two
effective parameters, $\Delta'=\Delta-U_0J_2N$, $U_0'=
U_0(J_1-J_2)$, and rewrite the Hamiltonian as
\begin{eqnarray}\label{hamiltonian redefined}
    H&=&-t(b_1^\dagger b_2+b_2^\dagger b_1)+\frac{u}{2}(n_1(n_1-1)+n_2(n_2-1)) \nonumber \\
    & &+ U_0' a^\dagger a n_1-\Delta'a^\dagger
a+\eta(a+a^\dagger),
\end{eqnarray}
then effectively we have $J_1=1$, $J_2=0$. Experimentally, excellent
control of the position of a single atom relative to the cavity mode
has been demonstrated, so atom-field coupling can be tailored as
wanted~\cite{kuhr, dotsenko, nufmann, chapman}. In the following, we
shall omit the prime for notational simplicity.

\section{analytical and Numerical analysis based on master equation}
\label{sec 3} The Hamiltonian derived above controls the coherent
evolution of the atom-field system. However, we still have to take
the dissipation into account, which comes from the cavity loss in
the model we consider. The overall evolution of the system is
governed by the master equation:
\begin{equation}\label{master eq}
   \dot{\rho}= -i[H,\rho]+\kappa(2a\rho a^\dagger- a^\dagger a \rho- \rho a^\dagger a)
   \equiv \mathcal {L} \rho.
\end{equation}
Here $\rho$ is the density matrix of the atom-field system in the
rotating frame, and $\kappa$ is the cavity loss rate. Note that
generally the frequency of the cavity mode falls in the optical
regime, hence the environment can be treated as at zero temperature.
The master equation will be our starting point for the rest of the
paper.

As for the dynamics of our system, we stress that there are two
distinct time scales \cite{meystre}. One is the inverse of the
atomic tunneling rate $t^{-1}$, the other being that of the cavity
loss rate $\kappa^{-1}$. They are the characteristic times of the
atomic and field subsystems, respectively. In typical experimental
situations, $\kappa$ is of order $10^6$ Hz, while $t$ (and $u$) is
of order $10^3$ Hz at most \cite{jaksch}. This means that generally
there is a hierarchy $t^{-1}\gg\kappa^{-1}$. The identification of
two different time scales leads us to classify the dynamics of the
system into \textsl{short-} and \textsl{long-} time behaviors, which
correspond to two disjoint time intervals, (\romannumeral 1)
$0<\tau\ll t^{-1}$ and (\romannumeral 2) $\tau\gg t^{-1}$,
respectively. In the short-time interval, the atomic tunneling is
``frozen''. However, we still expect the system to display some
non-trivial behaviors, because this time may be long in unit of
$\kappa^{-1}$. In the long-time interval, the atomic tunneling may
eventually give rise to some important results and should be taken
into full account. Specifically we divide the Hamiltonian $H$ into
tunneling and non-tunneling terms,
\begin{equation}\label{h=h_hop+h_non}
   H=H_{t}+H_{non},
\end{equation}
with
\begin{eqnarray}
  H_{t} &=& -t(b_1 ^\dagger b_2 +b_2 ^\dagger b_1),   \\
 H_{non} &=& -\Delta a^\dagger a+\eta(a+a^\dagger)+U_0 a^\dagger a n_1
  \nonumber \\
  &&+\frac{u}{2}(n_1(n_1-1)+n_2(n_2-1)),
\end{eqnarray}
and rewrite the master equation as
\begin{equation}\label{maser equa two parts}
    \dot{\rho}=-i[H_{non}, \rho]+\kappa(2 a \rho a^\dagger-a^\dagger a \rho- \rho a^\dagger
    a)-i[H_{t},\rho].
\end{equation}
The last term will be neglected (kept) in the short- (long-) time
intervals, respectively. In the following we shall investigate the
behavior of the system in the two time intervals both analytically
and numerically.
\subsection{Short-time behavior}
Let us assume initially the atoms are in the ground state $|G
\rangle$ of the Hamiltonian $H_a$, while the field is in the vacuum
state $|0\rangle_f$, i.e.,
\begin{equation}\label{initial state}
    \rho(0)=|G
\rangle \langle G| \otimes |0\rangle_{ff} \langle 0|.
\end{equation}
Then at $\tau=0$ the pump is turned on and the system evolves
according to the master equation (\ref{master eq}). In general,
solving a master equation analytically exactly is a formidable task,
so we will resort to numerical methods as we do. However, in the
short-time interval, as mentioned above, we may neglect the
tunneling term and approximate the master equation by
%\begin{eqnarray}\label{maser equa short2}
%    \dot{\rho}&=&-i[H_{non}, \rho]+\kappa(2 a \rho a^\dagger-a^\dagger a \rho- \rho a^\dagger
%    a)  \nonumber \\
%    &\equiv& \mathcal {L}_{non} \rho.
%\end{eqnarray}
\begin{equation}\label{master equa short}
      \dot{\rho}=-i[H_{non}, \rho]+\kappa(2 a \rho a^\dagger-a^\dagger a \rho- \rho a^\dagger
    a)
    \equiv \mathcal {L}_{non} \rho.
\end{equation}
As pointed out in Appendix A, $H_{non}$ can be mapped into the Dicke
model in the dispersive regime, up to some minor differences. The
dynamics of the Dicke model in a driven and damped cavity, in the
dispersive regime, has been studied in detail in Ref.~\cite{klimov}.
Here we shall follow the techniques there.

Under the transformation to another reference frame,
\begin{eqnarray}\label{trans}
     \tilde{\rho}=e^{iA\tau}\rho e^{-iA\tau}, \quad\quad\quad\quad  \\
    A=\frac{u}{2}(n_1(n_1-1)+n_2(n_2-1)),
\end{eqnarray}
the master equation (\ref{master equa short}) takes the form
\begin{equation}\label{maser equa short2}
    \dot{\tilde{\rho}}=-i[\tilde{H}, \tilde{\rho}]+\kappa(2 a \tilde{\rho} a^\dagger-a^\dagger a \tilde{\rho}- \tilde{\rho} a^\dagger
    a),
\end{equation}
with the simplified Hamiltonian
\begin{equation}\label{H eff}
    \tilde{H}=-\Delta a^\dagger a+\eta(a+a^\dagger)+U_0 a^\dagger a
    n_1.
\end{equation}
Note that $\tilde{H}$ is diagonal in the atomic space. This leads us
to expand the density matrix $\tilde{\rho}$ as
\begin{equation}\label{tilde rho}
    \tilde{\rho}=\sum_{m,n=0}^{N} |m\rangle\langle n | \otimes
    \tilde{\rho}_{mn},
\end{equation}
where $|m\rangle\equiv|m, N-m\rangle$ denotes the atomic state with
$m$ atoms in the left trap and $(N-m)$ atoms in the ritht trap, and
$\tilde{\rho}_{mn}=\langle m |\tilde{\rho} |n \rangle$, which is
still an operater in the field space. The initial values of the
$\tilde{\rho}_{mn}$'s are
\begin{equation}\label{rho mn}
    \tilde{\rho}_{mn}(0)=\langle m |G \rangle \langle G| n \rangle \cdot|
    0\rangle_{ff}\langle 0 |.
\end{equation}
In terms of $\tilde{\rho}_{mn}$, the atomic and field density
operators are respectively,
\begin{eqnarray}\label{field operator}
  \tilde{\rho}_a &=& tr_{f}(\tilde{\rho})=\sum_{m,n=0}^{N}tr_f(\tilde{\rho}_{mn}) |m\rangle\langle n |
    ,  \\
  \tilde{\rho}_f  &=& tr_a(\tilde{\rho})=\sum_{m=0}^N
  \tilde{\rho}_{mm}.
\end{eqnarray}
It is straightforward to obtain the time evolution equation of the
operators $\tilde{\rho}_{mn}$ from the master equation (\ref{maser
equa short2}),
\begin{eqnarray}{\label{motion of rho mn}}
  \dot{\tilde{\rho}}_{mn}&=& -i[-(\Delta-U_0 p)a^\dagger a+\eta(a+a^\dagger),\tilde{\rho}_{mn}]
  \nonumber    \\
   && +\kappa(2 a \tilde{\rho}_{mn} a^\dagger-a^\dagger a \tilde{\rho}_{mn}- \tilde{\rho}_{mn} a^\dagger
    a)           \nonumber  \\
    && -iU_0 q (a^\dagger a \tilde{\rho}_{mn}+ \tilde{\rho}_{mn} a^\dagger
    a),
\end{eqnarray}
where $p=(m+n)/2$, $q=(m-n)/2$. The general solution of this
equation is derived in Ref.~\cite{klimov} by applying the dynamical
symmetry method. The result is thorough but complicated, so we will
cite it only when we have to.

For the diagonal cases with $m=n$, $q=0$, Eq.~(\ref{motion of rho
mn}) reduces to the master equation describing the dynamics of a
single field mode subjected to damping and pumping. Up to a constant
coefficient, we make the ansatz that the solution of
Eq.~(\ref{motion of rho mn}) in this case takes the form
\begin{equation}\label{ansatz}
    \tilde{\rho}_{mm}(\tau)=| \alpha _m(\tau)\rangle _{ff}\langle \alpha _m(\tau)
    |,
\end{equation}
where $| \alpha _m(\tau)\rangle_f$ denotes the field coherent state,
and $ \alpha _m(0)=0$. Substituting (\ref{ansatz}) into
Eq.~(\ref{motion of rho mn}), after some operator manipulations, we
find that the equation is satisfied if
\begin{equation}\label{alpha motion}
    \dot{\alpha}_m(\tau)=-(\kappa-i(\Delta-U_0m))\alpha_m(\tau)-i\eta.
\end{equation}
This equation is readily solved by
\begin{equation}\label{alpha solution}
    \alpha_m(\tau)=\alpha_m(\infty)(1-e^{-(\kappa-i(\Delta-U_0m))\tau}),
\end{equation}
with
\begin{equation}\label{alpha m}
    \alpha_m(\infty)=\frac{-i\eta}{\kappa-i(\Delta-U_0m)}.
\end{equation}
Recalling Eqs.~(\ref{rho mn}) and ($27$), we have the field density
operator
\begin{equation}\label{field solution}
    \tilde{\rho}_f(\tau)=\sum_{m=0}^N |\langle m|G\rangle|^2 | \alpha _m(\tau)\rangle _{ff}\langle \alpha _m(\tau)
    |,
\end{equation}
which is an incoherent superposition of a series of coherent states.
Due to the atom-field coupling, both the weights of the coherent
states and the coherent states themselves, depend on the initial
atomic state. For times $\tau \gg \kappa^{-1}$, $\alpha _m(t)$
saturates to the value $\alpha_m(\infty)$ and the field approaches
the quasi-steady state
\begin{equation}\label{field quasi st}
    \tilde{\rho}_f(\infty)=\sum_{m=0}^N|\langle m|G\rangle|^2 | \alpha _m(\infty)\rangle _{ff}\langle \alpha _m(\infty)
    |.
\end{equation}
The average photon number in the quasi-steady state is
\begin{equation}\label{photon number st}
    \langle a^\dagger a\rangle= \sum_{m=0}^N|\langle m|G\rangle|^2 \frac{\eta^2}{\kappa^2+(\Delta-U_0m)^2}
    .
\end{equation}
Note that the field approaches its quasi-steady state in a time of
order $\kappa^{-1}$, which is well within the short-time interval
$0<\tau\ll t^{-1}$. This indicates that the analysis above is
self-consistent. We refer to this ``steady state'' of the field as
quasi-steady state so as to differentiate it from the true steady
state in the long-time interval.
\begin{figure}[t] \centering
\includegraphics[bb=10 25 312 215, width=0.5\textwidth]{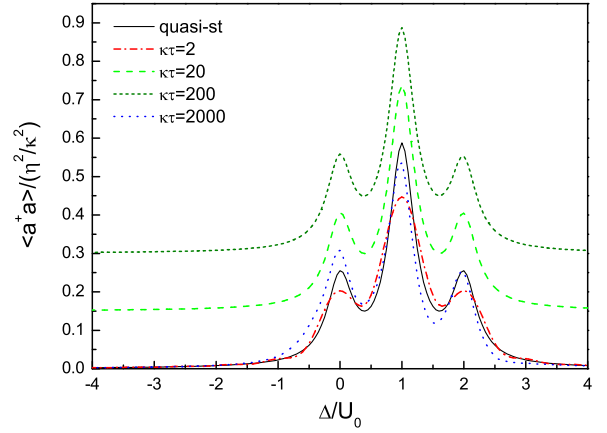}
\caption{(color online). Normalized photon number $\langle a^\dagger
a\rangle/(\eta^2/\kappa^2)$ as a function of the pump-cavity
detuning $\Delta$, with the master equation cut off at four
different times. The quasi-steady state result (solid line) is shown
for comparison. The two lines corresponding to $\kappa\tau=(20,
200)$ have been up shifted 0.15 and 0.30 respectively, unless they
coincide with the solid line. The parameters are $(t,
u)=2\pi\times(400, 200)$Hz, $(\kappa, U_0, \eta)=2\pi\times(1.5,
6.0, 0.1)\times10^6$Hz. The number of atoms is $N=2$.}\label{fig1}
\end{figure}

For off-diagonal cases with $m\neq n$, the last term in
Eq.~(\ref{motion of rho mn}) is nonzero. As pointed out in
Ref.~\cite{klimov, nemes}, this non-unitary term will result in the
complete disappearance of the operators $\tilde{\rho}_{mn}$, that
is, the complete coherence loss of the atomic subsystem. Explicitly,
\begin{equation}\label{decay of rho mn}
   |\rho^{mn}_a|=|tr_f(\tilde{\rho}_{mn})| \propto exp(-\tau/\tau_{mn}),
\end{equation}
with the ($m$, $n$)-dependent characteristic time
\begin{equation}\label{tau mn}
    \tau_{mn}=\frac{[\kappa^2+(\Delta-U_0m)^2][\kappa^2+(\Delta-U_0n)^2]}{\kappa U_0^2\eta^2(m-n)^2}.
\end{equation}
If all the $\tau_{mn}$'s are much smaller than $t^{-1}$, then
eventually the atomic subsystem will reach a purely mixed state,
\begin{equation}\label{mixed atomic state}
    \rho_a(\infty)=\sum_{m=0}^N |\langle m|G\rangle|^2 | m \rangle
    \langle m |,
\end{equation}
and the atom-field system is merely classically correlated,
\begin{equation}\label{rho classically correlated}
    \rho(\infty)= \sum_{m=0}^N |\langle m|G\rangle|^2 | m \rangle
    \langle m | \otimes |\alpha _m(\infty)\rangle _{ff}\langle \alpha _m(\infty)
    |.
\end{equation}
\begin{figure}[t]\centering
\includegraphics[bb=10 25 312 215, width=0.5\textwidth]{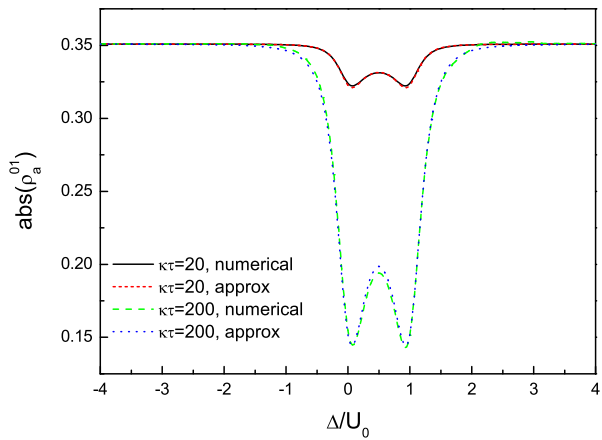}
\caption{(color online). Decay of the off-diagonal element
$\rho^{01}_a$ at two time sections, $\kappa\tau=(20, 200)$.
Analytical approximate results according to Eqs.~(\ref{decay of rho
mn}) and (\ref{tau mn}) and numerical results based on master
equation (\ref{master eq}) are shown for comparison. The parameters
are the same as in Fig.~\ref{fig1}.}\label{fig2}
\end{figure}

All the results derived above are based on the approximation that
the atomic tunneling is negligible in the short time interval. The
quality of this approximation is well demonstrated in
Figs.~\ref{fig1} and \ref{fig2}. There we show the time evolution of
the photon number and off-diagonal element $\rho_a^{01}$. The
results are obtained by numerically integrating the master equation
(\ref{master eq}) with the atomic tunneling being taken into
account. As shown in Fig.~\ref{fig1}, within a time of order
$\kappa^{-1}$, the photon number builds up and saturates to the
value given by Eq.~(\ref{photon number st}). Then it holds on to
times of order $10^3/\kappa$ before signatures of deviation from the
approximation arise. The excellent agreement between the analytical
and numerical results is again demonstrated in the decay of the
off-diagonal element $\rho_a^{01}$ in Fig.~\ref{fig2}.
\subsection{Long-time behavior}
\begin{figure}[t]
\centering
\includegraphics[bb=10 25 312 215, width=0.5\textwidth]{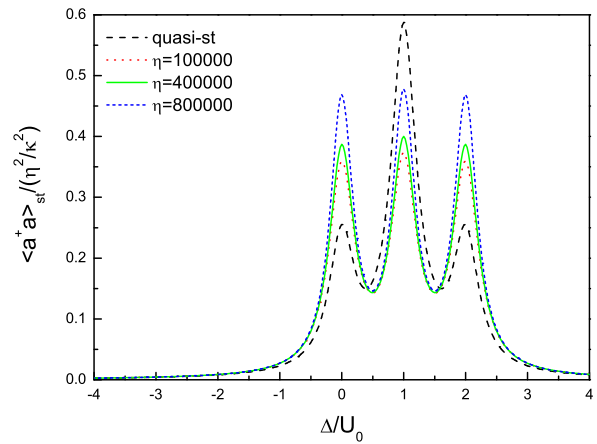}
\caption{(color online). Normalized photon number at steady state
$\langle a^\dagger a\rangle _{st}/ (\eta^2/\kappa^2)$ as a function
of the pump-cavity detuning $\Delta$ for three different pump
strengthes. The dashed line corresponding to the quasi-steady state
results given by Eq.(\ref{photon number st}) is shown for
comparison. The parameters are the same as in Fig.~\ref{fig1}.}
\label{fig3}
\end{figure}

\begin{figure}[t]
\centering
\includegraphics[bb=10 25 312 215, width=0.46\textwidth]{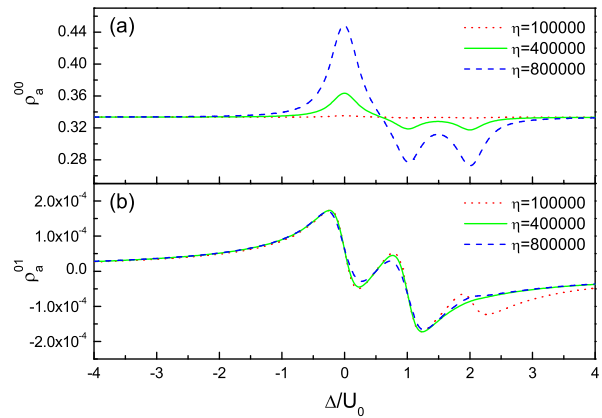}
\caption{(color online). (a) Diagonal element $\rho_a^{00}$ and (b)
off-diagonal element $\rho_a^{01}$ of the reduced atomic density
matrix $\rho_a$ at steady state. The parameters are the same as in
Fig.~\ref{fig1}.} \label{fig4}
\end{figure}

As shown above, in the short-time interval, the atomic tunneling can
be neglected. However, in the long-time interval, where the system
enters the steady state $\rho_{st}$, the atomic tunneling does play
an important role. An analytical exact solution of $\rho_{st}$ is
unavailable, so we rely on numerical methods \cite{tan}. In
Fig.~\ref{fig3} we show the the normalized photon number in steady
state as a function of the detuning $\Delta$ with the pump strength
varied. The difference between the long-time steady state results
and the short-time quasi-steady state result is apparent. A striking
feature of the spectrum in steady state is that the peaks are almost
of equal height and in particular, in the weak pump limit
($\eta/\kappa \ll 1$), the height converges to some value around
$1/3$ (take into account the overlap between the peaks). In
contrast, for the specific set of parameters in our numerical
calculations, $(t,u)=2\pi\times(400, 200)$ Hz and $N=2$, the
quasi-steady state result Eq.~(\ref{photon number st}) predicts the
heights of the three peaks to be $0.23$, $0.53$ and $0.23$,
respectively. The difference between the steady state and
quasi-steady state may be more directly revealed in Fig.~\ref{fig4},
where we present the diagonal element $\rho_{a}^{00}$ and
off-diagonal element $\rho_{a}^{01}$ of the atomic density matrix
$\rho_{a}$ as the detuning and pump strength are varied. From
Fig.~\ref{fig4}(a) we see that when the detuning is far from all
possible resonances, the element $\rho_{a}^{00}$ is around $1/3$
regardless of the pump strength; and in the limit of weak pump,
$\rho_{a}^{00}$ is around $1/3$ in the whole range of the detuning.
From Fig.~\ref{fig4}(b) we see the off-diagonal element
$\rho_{a}^{01}$ is far less than unity in the domain of $\Delta$ and
$\eta$ we consider. Other diagonal and off-diagonal elements have
similar behavior and hence are not shown.

We also investigated the cases with $N\neq2$, and some common
features are found. That is, as long as the condition $(t,
u)\ll(\kappa, U_0)$ is fulfilled, in the weak pump limit the
normalized photon number in steady state $\langle a^\dagger a\rangle
_{st}/ (\eta^2/\kappa^2)$ as a function of the detuning $\Delta$ is
the superposition of $N+1$ lorentzians, which are centered at
$\Delta=U_0s$ $(s=0,1,\ldots N)$, and of heights nearly $1/(N+1)$.
Besides that, the diagonal elements of the atomic density matrix
converge to values around $1/(N+1)$, while all the off-diagonal
elements are vanishingly small, i.e., the atomic subsystem is in a
nearly absolutely ``unpolarized'' mixed state.

The analysis of the short-time behaviors can in fact help us to
understand the features of the steady state. The steady state
$\rho_{st}$ satisfies
\begin{equation}\label{master st}
    0=\mathcal {L} \rho_{st}=\mathcal {L}_{non} \rho_{st}-i[H_t,\rho_{st}].
\end{equation}
Because $t\ll(\kappa, U_0)$, we shall treat the second term as a
perturbation over the first term, for which we have analytical
results. Assume that
\begin{equation}\label{rho=rho0+rho1}
    \rho_{st}=\rho_{st}^0+\rho_{st}^1,
\end{equation}
where $\rho_{st}^0$ is of zeroth order in $t/\kappa$, while
components of higher orders in $t/\kappa$ are included in
$\rho_{st}^1$. $\rho_{st}^0$ satisfies the equation $\mathcal
{L}_{non} \rho_{st}^0=0$. According to the analysis in the previous
subsection, its general solution is
\begin{equation}\label{solution for rho0}
   \rho_{st}^0=\sum_{m=0}^{N} C_m |m\rangle
\langle  m| \otimes |\alpha_m(\infty)\rangle \langle
\alpha_m(\infty)|,
\end{equation}
with the coefficients $C_m$ being arbitrary. Note that $\rho_{st}^0$
is diagonal in the atomic space, which implies that the off-diagonal
elements of the atomic density matrix must come from $\rho_{st}^1$
and hence are at least of order $t/\kappa$. This explains why the
off-diagonal elements are vanishingly small as revealed by the
numerical calculations. The physical picture is that, via the
atom-field coupling and the dissipation, the coherence of the atomic
subsystem is greatly depleted, the remaining weak coherence is just
due to the finite atomic tunneling.

The knowledge of the off-diagonal elements of the atomic density
matrix allows us to understand the behavior of the diagonal elements
and the photon number, at least in the weak pump limit. In steady
state, we have the following equation for an arbitrary operator
$\hat{O}$,
\begin{equation}\label{expectation equation under steady }
    0=\langle \dot{\hat{O}}\rangle_{st}=-i\big\langle[ \hat{O},H
    ]\big\rangle_{st} +\kappa \big\langle [a^\dagger, \hat{O}]a-a^\dagger [a,\hat{O}]
    \big\rangle_{st}.
\end{equation}
Let $\hat{O}=|m\rangle \langle  m+1|$, $0\leq m \leq N-1$, then we
obtain
\begin{eqnarray}
0&=&f(m+1)\big(\langle  |m\rangle \langle m| \rangle_{st}-\langle
|m+1\rangle \langle m+1|
    \rangle_{st}\big) \quad\quad\quad\quad \nonumber \\
    &&  +f(m+2)\langle  |m\rangle \langle  m+2|
    \rangle_{st}-f(m)\langle  |m-1\rangle \langle  m+1|
    \rangle_{st} \nonumber \\
    &&-\frac{u}{t}(2m+1-N)\langle  |m\rangle \langle  m+1|
    \rangle_{st}   \nonumber \\
    &&-\frac{U_0}{t}\langle  |m\rangle \langle m+1|a^\dagger a
    \rangle_{st},
\end{eqnarray}
where $f(m)=\sqrt{(m+1)(N-m)}$. In the weak pump limit, the second
and third terms on the right hand side is of order $t/\kappa$, the
fourth term $u/\kappa$, while the fifth term $(\eta/\kappa)^2$, so
to zeroth order in $t/\kappa$, $u/\kappa$ and $\eta/\kappa$, we have
\begin{equation}\label{equal}
    \langle  |m\rangle \langle  m| \rangle_{st}-\langle  |m+1\rangle \langle  m+1|
    \rangle_{st}=0.
\end{equation}
This equation, together with the normalization condition
$tr(\rho_a)=1$, means that to zeroth order in $t/\kappa$, $u/\kappa$
and $\eta/\kappa$,
\begin{equation} \label{m+1}
\langle
   |m\rangle
\langle  m|\rangle_{st}= \frac{1}{N+1}.
\end{equation}
Returning to Eqs.~(\ref{rho=rho0+rho1}) and (\ref{solution for
rho0}), we see that in the weak pump limit (the condition $t, u\ll
\kappa$ is spontaneously satisfied), the steady state is well
approximated by
\begin{equation}\label{appr rho by rho0}
    \rho_{st}\simeq\frac{1}{N+1}\sum_{m=0}^N |m\rangle \langle
    m|\otimes |\alpha_m(\infty)\rangle_{ff} \langle  \alpha_m(\infty)|.
\end{equation}
The photon number $\langle a^\dagger a\rangle_{st}$ in this limit is
given by
\begin{equation}\label{photon number steady}
    \langle a^\dagger a\rangle_{st}=\frac{1}{N+1}\sum_{m=0}^N \frac{\eta^2}{\kappa^2+(\Delta-U_0
    m)^2}.
\end{equation}
Eqs.~(\ref{m+1}) and (\ref{photon number steady}) account for the
weak pump steady state features. In fact the photon number is always
directly determined the diagonal elements $\langle |m\rangle \langle
m| \rangle_{st}$, not limited to the weak pump limit. Let
$\hat{O}=a^\dagger a$ and $a|m\rangle \langle m|$ in
Eq.~(\ref{expectation equation under steady }), we have
\begin{eqnarray}
\label{algebra}
  0&=&i\eta(\langle a \rangle_{st} -\langle a^\dagger \rangle_{st})-2\kappa\langle a^\dagger
    a\rangle_{st}     \nonumber \\
   &=& i\eta \sum_{m=0}^N\big(\langle a|m\rangle \langle m|
   \rangle_{st}-c.c.
   \big)-2\kappa\langle a^\dagger
    a\rangle_{st}, \label{ata motion}   \\
  0 &=& it\big\langle[a|m\rangle \langle m|, b_1^\dagger b_2+b_2^\dagger b_1]\big\rangle_{st}-i\eta\left\langle |m\rangle \langle m|\right\rangle_{st} \nonumber\\
   && -\big(\kappa-i(\Delta -U_0m)\big)
   \langle
a|m\rangle \langle m| \rangle_{st},      \label{amm motion}
\end{eqnarray}
where $c.c.$ stands for complex conjugate. As shown in Appendix B,
the first term on the right hand side of Eq.~(50) can be safely
neglected compared to the second term. By neglecting it we get a set
of algebraic equations of $\langle a^\dagger
    a\rangle_{st}$, $\langle
a|m\rangle \langle m| \rangle_{st}$. We solve
\begin{equation}\label{photon number at st}
    \langle a^\dagger a\rangle_{st}\simeq \sum_{m=0}^{N} \langle
   |m\rangle
\langle  m|\rangle_{st}\cdot\frac{\eta^2}{\kappa^2+(\Delta
-U_0m)^2},
\end{equation}
which is valid for arbitrary values of $\eta$.

\begin{figure}[t]
\centering
\includegraphics[bb=10 25 322 225, width=0.5\textwidth, height=5cm]{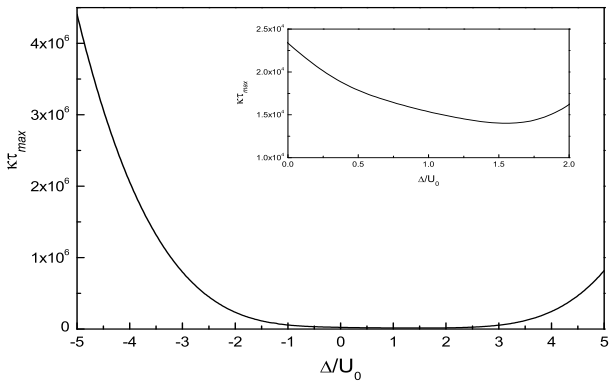}
\caption{(color online). Timescale $\tau_{max}$ in unit of
$1/\kappa$ for the system approaching the steady state. Inset:
close-up of the curve in the regime $0\leq \Delta/U_0 \leq 2$. The
parameters are the same as in Fig.~\ref{fig1}.} \label{fig5}
\end{figure}

An important question of concern is what is the time scale for the
system to approach the steady state. The general solution of a
master equation like Eq.(\ref{master eq}) with a time-independent
Liouvillian can be written as a sum of a series of complex
exponentials,
\begin{equation}\label{exponential series}
    \rho(\tau)=\sum_j a_j exp(s_j \tau),
\end{equation}
where $s_j=-R_j+iI_j$, ($R_j, I_j\in \mathcal {R}$) are the
eigenvalues of the Liouvillian, while the coefficients $a_j$ are
determined by the initial conditions. As well known, the Liouvillian
is singular and has at least one zero eigenvalue which correspond(s)
to the steady state(s), and all the non-zero eigenvalues have
negative real parts. This ensures that $\rho(\tau)$ converges to the
steady state(s) in the limit of $\tau\rightarrow \infty$. Obviously,
the time scale of this process is set by the inverse of the least
modulus real part of the eigenvalues. We define
\begin{equation}\label{time scale}
    \tau_{max}=\max_j\{\frac{1}{R_j}; R_j \neq 0\}.
\end{equation}
In Fig.\ref{fig5}, we show $\kappa \tau_{max}$ as a function of the
detuning. Note that $\kappa \tau_{max}$ is of order $10^4\sim10^6$
in the regime $|\Delta/U_0|\leq 5$ and it diverges as
$|\Delta|\rightarrow \infty$. Compared with Fig.~\ref{fig1}, this
clearly demonstrates that the long-time and short-time behaviors lie
in two well separated time intervals. Moreover, it clarifies a point
that, when the pump is far from all possible resonances, the
influence of the pump on the system is negligible, so it would take
the system a time experimentally unaccessible to reach the steady
state. Of course, the situation is still good in the regime $0\leq
\Delta/U_0 \leq 2$, where the time scale is of order $10$ms. At
present, individual atoms have been trapped and detected in an
optical cavity for time scales exceeding $15$s \cite{chapman}, so we
expect that the steady state features may also have the possibility
to be observed in the future.
\section{Summary and remarks}
\label{sec 4} We investigated the dynamics of a dispersively
interacting atom-field system, with the slowly varying atomic
interwell tunneling coupled with the fast varying field dynamics.
Depending on the role of the atomic tunneling, the dynamics of the
system was classified into short-time and long-time behaviors.

In the short-time interval ($0<\tau\ll t^{-1}$), as verified by the
numerical calculations, the atomic tunneling can be neglected. We
recovered the result of Ref.~\cite{Ritsch_nature} in the
``two-site'' case, and went beyond to obtain a more detailed picture
of the dynamics of the atom-field system, such as the decoherence of
the atomic subsystem, the correlation between the atomic and field
subsystems. In our analysis, a central observation is the analogy
between the model we consider and the well known Dicke model in the
dispersive regime. In fact, many results are directly borrowed from
previous works on Dicke model \cite{klimov}. Of course, we stress
that this similarity is not essential. It is the dispersive nature
of the atom-field coupling that counts. As can be seen from our
procedures, similar techniques and results apply also to the
many-site cases, e.g., the original model in
Ref.~\cite{Ritsch_nature}.

As for the long-time behavior, we were primarily interested in the
steady state. If the atomic tunneling is absent, the steady state of
the system is in the form of Eq.~(\ref{solution for rho0}). The
atomic and field subsystems are only classically correlated, and the
populations of different atomic states are absolutely determined by
the initial state. However, the presence of atomic tunneling leads
to strong population transfers between the atomic states. A
remarkable feature is that, in the weak pump limit, the atomic
states are almost equally populated, which is substantially
different from the ground state atomic distribution. We also
quantitatively investigated the time scale of reaching the steady
state and found that it lies well in the long time interval and is
accessible under the experimental situations at present.

Finally, we have some remarks about the experimental implementation
of the model we discussed in this work. The double-well may be
constructed by two adjacent optical dipole traps between which the
distance can be adjusted as in Ref.~\cite{grangier}. An important
feature of these traps is the extremely small focal spot, a beam
waist radius of $w_0 <1 \mu m$ is achieved. This helps to confine
the atom in a very small volume and validate the ``tight confinement
approximation''. The case $J_1=1$, $J_2=0$ occurs when one atomic
mode is localized in an antinode of the field mode, and the other in
a node.

\begin{acknowledgments}
We are grateful to P. Zhang and M. Xiao for stimulating discussions
and we would also like to thank A.~S. Parkins for his help on the
programming. This work was supported by NSF of China under grant
90406017, 60525417, 10775176, the NKBRSF of China under Grant
2005CB724508, 2006CB921400, 2006CB921206, and 2006AA06Z104..

\end{acknowledgments}

\appendix
\section{connection with the Dicke Model}
In terms of Schwinger's representation of the angular momentum
operators \cite{sakurai},
\begin{eqnarray}
  S_x &=& \frac{1}{2}(b_2^\dagger b_1 +b_1^\dagger b_2), \\
  S_y &=& \frac{i}{2}(b_2^\dagger b_1 -b_1^\dagger b_2), \\
  S_z &=& \frac{1}{2}(b_1^\dagger b_1 -b_2^\dagger b_2),
\end{eqnarray}
the Hamiltonian $H$ can be rewritten as
\begin{equation}\label{h=h-hop+h-non}
    H=H_{t}+H_{non},
\end{equation}
with
\begin{eqnarray}
  H_{t} &=& -2t S_x, \\
  H_{non} &=& (\frac{U_0N}{2}-\Delta)a^\dagger a+\frac{U_0}{2}(2a^\dagger a+1)S_z+\eta(a+a^\dagger) \nonumber\\
   && +u(S_z^2+\frac{N^2}{4}-\frac{N}{2})-\frac{U_0}{2}S_z.
\end{eqnarray}
Up to terms diagonal in the $S_z$-representation, $H_{non}$
corresponds to the Dicke model in the dispersive regime
\cite{klimov}, with cavity-pump detuning $(\frac{U_0N}{2}-\Delta)$,
effective atom-field coupling $\frac{U_0}{2}$ and pump strength
$\eta$. It is the two center-of-mass motion modes that correspond to
the two atomic internal levels involved in the Dicke model.

In this formalism, it is clear that the role of $H_{t}$ is to induce
transitions between different eigenstates of $S_z$ (that is, the
$|m\rangle$'s, $S_z|m\rangle=(m-\frac{N}{2})|m\rangle$), with
amplitudes proportional to $t$. However, since $t\ll \frac{U_0}{2},
\kappa$, this process can be neglected in the short-time interval.
\section{some useful inequalities}
We first introduce an inequality \cite{puri},
\begin{equation}\label{general inequality}
      \left|\langle A^\dagger
    B\rangle\right|^2 \leq\langle A^\dagger A\rangle \langle B^\dagger B\rangle,
\end{equation}
where $A$, $B$ are two arbitrary operators (hermitian or
non-hermitian), and the average is taken over an arbitrary state
(pure or mixed). Let $A=a$, $B=I$ (unity operator) in
Eq.(\ref{general inequality}), we obtain
\begin{equation}\label{inequality2}
  \langle a^\dagger   \rangle \langle  a
   \rangle \leq \langle a^\dagger a  \rangle  .
\end{equation}
For steady states, we have this inequality plus the constraint
Eq.(\ref{ata motion}), hence
\begin{eqnarray}
   \langle a^\dagger a  \rangle_{st} & \geq & \frac{1}{4}\big[(\langle a^\dagger\rangle_{st}+\langle a\rangle_{st})^2-(\langle a^\dagger\rangle_{st}-\langle
   a\rangle_{st})^2\big]
   \nonumber \\
    &=& \frac{1}{4}(\langle a^\dagger\rangle_{st}+\langle a\rangle_{st})^2+\frac{\kappa^2}{\eta^2}\langle a^\dagger
    a\rangle_{st}^2 \nonumber \\
    &\geq& \frac{\kappa^2}{\eta^2}\langle a^\dagger a\rangle_{st}^2,
\end{eqnarray}
which yields an upper bound for the photon number
\begin{equation}\label{upper bound}
     \langle a^\dagger a\rangle_{st} \leq \frac{\eta ^2}{\kappa ^2}.
\end{equation}
This can be understood as the maximum photon number occurs when the
probe is at resonance with the cavity.

We then show that it is legitimate to neglect the first term on the
right hand side of Eq.(\ref{amm motion}), which is
\begin{eqnarray}
\label{os hop}
  && \quad it\big\langle[a|m\rangle \langle m|, b_1^\dagger b_2+b_2^\dagger b_1]\big\rangle_{st}\quad  \nonumber   \\
  &&= it f(m+1)\big\langle a|m\rangle
\langle m+1|\rangle_{st}-it f(m+1)\langle a|m+1\rangle
\langle m|\big\rangle_{st}     \nonumber \\
   &&  \quad +it f(m)\big\langle a|m\rangle
\langle m-1|\rangle_{st}-it f(m)\langle a|m-1\rangle \langle
m|\big\rangle_{st}.
\end{eqnarray}
It is ready to show that the ratios of the four terms to
$-i\eta\langle |m\rangle \langle m|\rangle_{st}$ are at least of
order $t/\kappa$, as long as $\langle |m\rangle \langle
m|\rangle_{st}$ is of order unity. Let us take the third term for an
example. Applying inequalities (\ref{general inequality}) and
(\ref{upper bound}),
\begin{eqnarray}
  |it f(m) \langle a |m\rangle
\langle m-1| \rangle_{st}| &\leq& t f(m)\big [\langle a^\dagger
a\rangle_{st} \langle|m\rangle
\langle m| \rangle_{st}\big]^{1/2} \nonumber \\
  &\leq& t f(m) \langle a^\dagger a \rangle_{st}^{1/2} \nonumber \\
  &\leq& f(m)(\frac{t}{\kappa}) \eta,
\end{eqnarray}
and similar results hold for the other three terms. This guarantees
that the first term on the right hand of Eq.(\ref{amm motion}) is
much less than the second term.


\begin{thebibliography}{60}
\bibitem{esslinger nature}
F. Brennecke, T. Donner, S. Ritter, T. Bourdel, M. K$\ddot{o}$hl,
and T. Esslinger, Nature (London) (in press).

\bibitem{tonature}
Y. Colombe, T. Steinmetz, G. Dubois, F. Linke, D. Hunger, and J.
Reichel, Preprint at http://cn.arxiv.org /abs/0706.1390.

\bibitem{chapman}
K.~M. Fortier, S.~Y. Kim, M.~J. Gibbons, P. Ahmadi, and M.~S.
Chapman, Phys. Rev. Lett. {\bf 98}, 233601 (2007).

\bibitem{nufmann} S. Nu{\ss}mann, M. Hijlkema, B. Weber, F. Rohde, G. Rempe, and A. Kuhn, Phys. Rev. Lett. {\bf
95}, 173602 (2005).

\bibitem{Ritsch_nature}
I.~B. Mekhov, C. Maschler, and H. Ritsch, Nature Physics. {\bf 3},
319 (2007).

\bibitem{ye}
J. Ye, D.~W. Vernooy, and H.~J. kimble, Phys. Rev. Lett. {\bf 83},
4987 (1999).

\bibitem{esslinger}
A. $\ddot{O}$ttl, S. Ritter, M. K$\ddot{o}$hl, and T. Esslinger,
Phys. Rev. Lett. {\bf95}, 090404 (2005); T. Bourdel, T. Donner, S.
Ritter, A. $\ddot{O}$ttl, M. K$\ddot{o}$hl, and T. Esslinger, Phys.
Rev. A {\bf 73}, 043602 (2006); S. Ritter, A. $\ddot{O}$ttl, T.
Donner, T. Bourdel, M. K$\ddot{o}$hl, and T. Esslinger, Phys. Rev.
Lett. {\bf 98}, 090402 (2007).


\bibitem{optics comm}
C. Maschler and H. Ritsch, Opt. Commun. {\bf243}, 145 (2004).

\bibitem{klimov} S.~M. Chumakov, A.~B. Klimov, and C. Saavedra, Phys. Rev. A {\bf 61},
033814 (2000).

\bibitem{ritsch prl}
C. Maschler and H. Ritsch, Phys. Rev. Lett. {\bf 95}, 260401 (2005).

\bibitem{meystre}
W. Chen, D. Meiser, and P. Meystre, Phys. Rev. A {\bf75}, 023812
(2007).

\bibitem{d f walls}
G.~J. Milburn, J. Corney, E.~M. Wright, and D.~F. Walls, Phys. Rev.
A {\bf 55}, 4318 (1997).

\bibitem{jaksch}
D. Jacksch, C. Bruder, J.~I. Cirac, C.~W. Gardiner, and P. Zoller,
Phys. Rev. Lett. {\bf81}, 3108 (1998).

\bibitem{stanper kurn}
S. Gupta, K.~L. Moore, K.~W. Murch, and D.~M. Stamper-Kurn, Preprint
at http://arxiv.org/abs/ 0706.1052 (2007).

\bibitem{kuhr} S. Kuhr, W. Alt, D. Schrader, M. M$\ddot{u}$ller, V. Gomer, and D. Meschede, Science {\bf 293}, 278
(2001).

\bibitem{dotsenko} I. Dotsenko, W. Alt, M. Khudaverdyan, S. Kuhr, D. Meschede, Y. Miroshnychenko, D. Schrader, and A. Rauschenbeutel, Phys. Rev. Lett. {\bf
95}, 033002 (2005).

\bibitem{nemes} J.~G. Peixoto de Faria and M.~C. Nemes, Phys. Rev. A
{\bf69}, 063812 (2004).

\bibitem{tan}
S.~M. Tan, J. Opt. B: Quantum Semiclass Opt. {\bf 1}, 424 (1999).

\bibitem{grangier}
J. Beugnon, C. Tuchendler, H. Marion, A. Ga$\ddot{e}$tan, Y.
Miroshnychenko, Y.~R.~P. Sortais, A.~M. Lance, M.~P.~A. Jones, G.
Messin, A. Browaeys, and P. Grangier, Nature Physics (published
online).

\bibitem{sakurai}
J.~J. Sakurai, Modern Quantum Mechanics, revised version
(Addison-Wesley, New York, 1994).

\bibitem{puri}
R.~R. Puri, Mathematical Methods of Quantum Optics (Springer,
Berlin, 2001).
\end{thebibliography}
\end{document}